\documentclass[runningheads]{llncs}
\usepackage{graphicx}

\usepackage{multirow} 
\usepackage{pdflscape} 
\usepackage{hyperref} 
\usepackage{xcolor} 
\usepackage{subcaption} 
\usepackage{amssymb} 
\usepackage{tcolorbox}
\usepackage{xspace}
\hypersetup{
    colorlinks=true,
    linkcolor=blue,
    filecolor=magenta,      
    urlcolor=blue,
    pdftitle={Pricings Report},
    pdfpagemode=FullScreen,
}

\newcommand{\system}{system\xspace}
\newcommand{\SaaS}{SaaS\xspace}

\newcounter{rqAnswerCounter}

\begin{document}

\title{Pricing4SaaS: Towards a pricing model to drive the operation of SaaS}

\author{Alejandro García-Fernández\inst{1}\orcidID{0009-0000-0353-8891} \and
José Antonio Parejo\inst{1}\orcidID{0000-0002-4708-4606} \and
Antonio Ruiz-Cortés\inst{1}\orcidID{0000-0001-9827-1834}}
\authorrunning{A. García-Fernández et al.}

\institute{SCORE Lab, I3US Institute, Universidad de Sevilla, Espa\~na \\
\email{\{agarcia29,japarejo,aruiz\}@us.es}}
\maketitle              
\begin{abstract}

The Software as a Service (SaaS) model is a distribution and licensing model that leverages pricing structures and subscriptions to profit. The utilization of such structures allows Information Systems (IS) to meet a diverse range of client needs, while offering improved flexibility and scalability. However, they increase the complexity of variability management, as pricings are influenced by business factors, like strategic decisions, market trends or technological advancements. In pursuit of realizing the vision of pricing-driven IS engineering, this paper introduces Pricing4SaaS as a first step, a generalized specification model for the pricing structures of systems that apply the Software as a Service (SaaS) licensing model. With its proven expressiveness, demonstrated through the representation of 16 distinct popular SaaS systems, Pricing4SaaS aims to become the cornerstone of pricing-driven IS engineering.


\keywords{Cloud-based IS engineering  \and Pricing \and Software as a Service}
\end{abstract}

\section{Introduction and Motivation}

The Software as a Service (SaaS) model is a distribution and licensing model wherein software is delivered as a service on a subscription basis. Instead of purchasing and installing software locally on premises, users access the software and its features through a web browser, or an Application Programming Interface (API), over the cloud \cite{cusumano2010cloud}. When an IS is provided as SaaS\footnote{For the sake of brevity, in this paper, the term SaaS and the expression Information Systems provided as SaaS will be used interchangeably.}, it is also a service, and as such ``is a means of delivering value to customers by facilitating outcomes customers want to achieve, without the ownership of specific costs and risks’’ \cite{itilv4}. 

The adoption of pricing plans (henceforth pricings), alongside subscription models, has emerged as the predominant licensing mechanism within these systems, as they allow to meet a diverse range of client needs, providing a predictable revenue stream for providers, and offering flexibility and scalability to users \cite{Jiang2009}. Such adoption has also been pivotal in the growth and sustainability of SaaS development, representing a shift from traditional software licensing to a more dynamic and user-centric approach.

Capacities collected in such structures are referred as ``features’’, which we define as ``distinctive characteristics whose presence/absence may guide a user’s decision towards a particular subscription’’. This notion provides a broader scope than the one given for Software Product Lines \cite{10.1007/11431855_34} or feature toggling \cite{FOWLER2023,jezequel2022feature,MAHDAVI2022}. In both cases, the concept tends to be fundamentally limited to increases in functionality \cite{IS2010}, leaving out aspects such as capacity or quality of service. 

Despite its advantages, anchoring the SaaS business model on dynamic pricing strategies means that market forces frequently shape the pricing structure (addition of features, alteration of usage limits, etc) \cite{LABPACK}. In order to sustain competitiveness and ensure that pricing plans are effectively met, those changes must be timely implemented by the developers and properly supported by the architecture and deployment infrastructure of the system. We coin this process —wherein business decisions influence pricing modifications, thereby initiating development efforts and changes on the architecture or deployment infrastructure— as the \emph{Pricing-driven development and operation of SaaS}.




Given this context, we consider that formalizing pricings is the foundational step towards offering support to these processes in a manner that is both cost-effective and efficient, laying a solid groundwork for the automating the analysis and management of the variability introduced by pricings within SaaS products. Although some researchers are already working on standardizing the definition of APIs' pricing \cite{PRICING4API} —a specific subset of SaaS— our primary goal is to design a general model that is able to represent the pricing for all varieties of SaaS, even though other models provide greater granularity for specific types of such systems.



From now on, after introducing our vision of \textit{Pricing-driven development and operation of SaaS} and the main objective of the paper, the following contributions are made: i) Pricing4SaaS as an specification model to represent pricings of SaaS (section \ref{section:pricing4SaaS}), and ii) Yaml4SaaS, a syntax that turns Pricing4SaaS available to be used by the community (section \ref{section:validation}), and for which we provide a validation tool \cite{YAML4SAASVALIDATOR}. Finally we conclude the paper in section \ref{section:conclussions} and propose future work within this research area.

\section{Modeling the pricing of SaaS}
\label{section:pricing4SaaS}

Although there is not yet a commonly accepted model that represents the pricing of SaaS, most providers share similar structures with multiple common elements. In this context, a pricing performs as a container of features, which relies on \emph{plans} and \emph{add-ons} to group them. The main difference between the latter two is that while users can only subscribe to one single plan, add-ons do not share such restriction, allowing users to subscribe to as many as they want. Within this model, a customer interacts with the SaaS by establishing \emph{a subscription}, through which he commits to pay a periodic fee (aka usage tariff \cite{LI2022}) to gain the ability to access and leverage the functionality and information provided by the SaaS in the terms and limits set out by the chosen plan or/and set of add-ons.

\begin{figure}[htb]
    \centering
    \includegraphics[width=\textwidth]{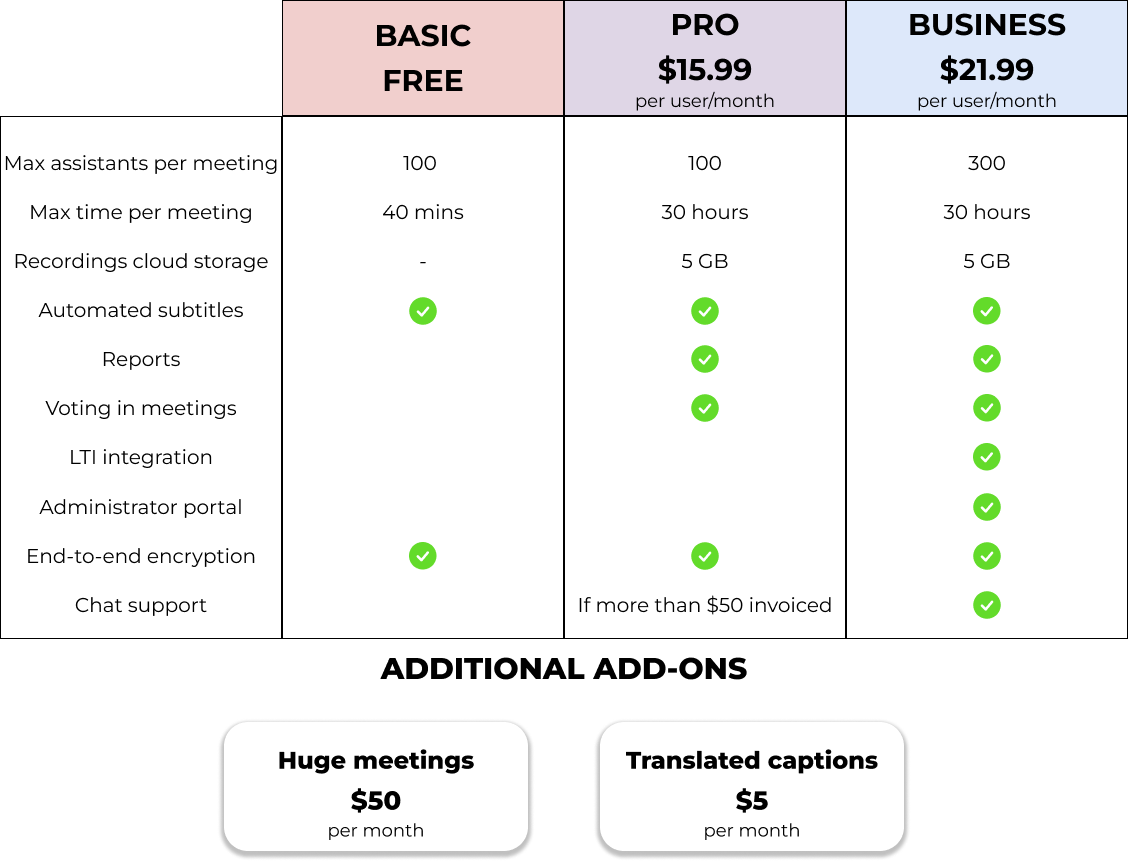}
    \caption{Partial view of Zoom's pricing}
    \label{fig:zoomPricing}
\end{figure}

To better explain the upcoming concepts, we will use Zoom as a running example, a cloud-based video conferencing service that enables users to virtually meet with others - either by video, audio-only or both, while conducting live chats - allowing users to record the sessions to view later. An adapted view of its pricing is represented in Figure \ref{fig:zoomPricing}, and contains a subset of features from the original one\footnote{Zoom's full pricing can be found \href{https://zoom.us/pricing}{here}}. In general, the pricing consists of three plans -which manage 10 features- and two add-ons -that manage one feature each-.

\subsection{The Pricing4SaaS model}

\begin{figure}
    \centering
    \makebox[\textwidth][c]{\includegraphics[width=\textwidth]{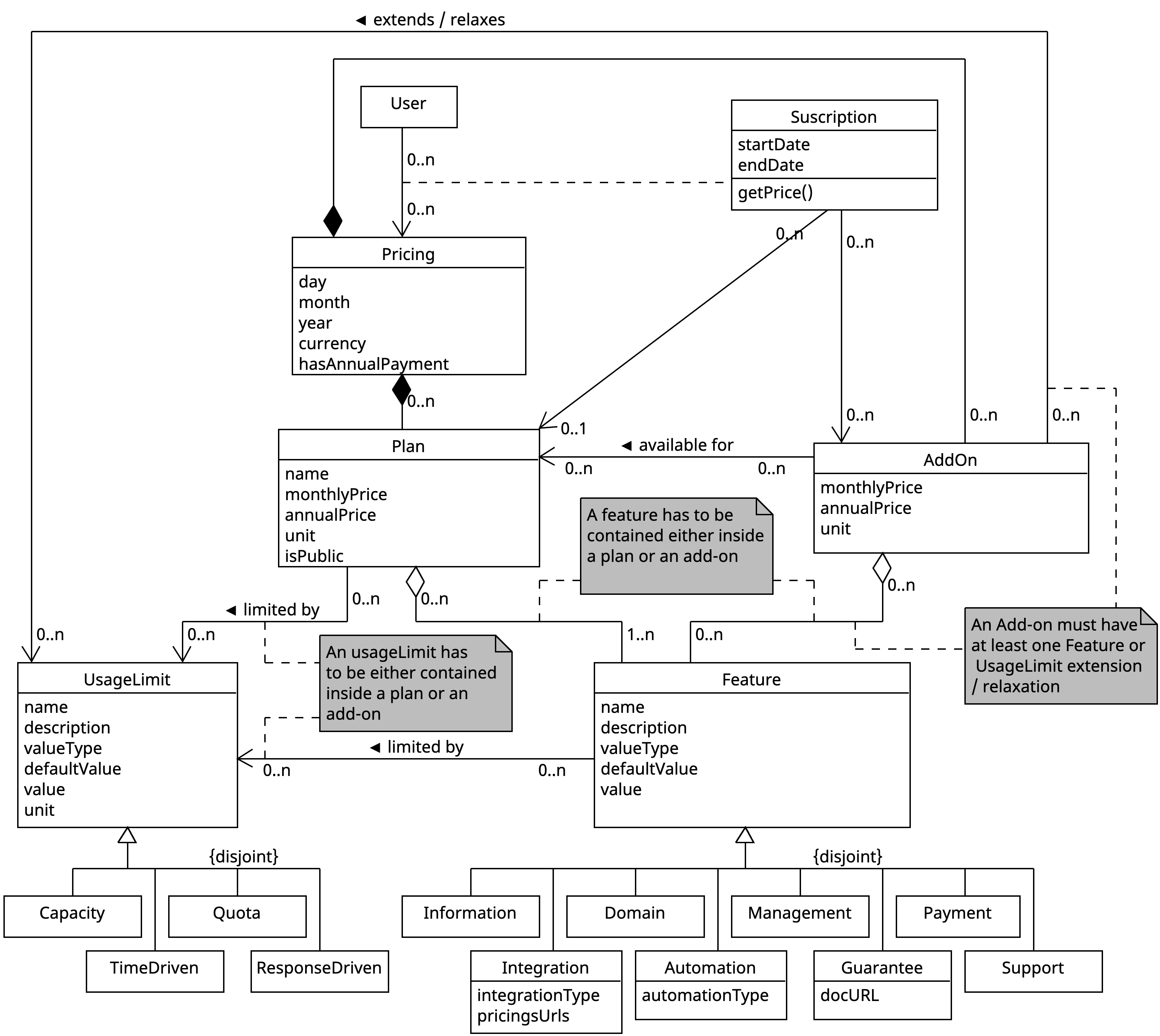}}
    \caption{Pricing4SaaS model}
    \label{fig:umlModel}
\end{figure}

Following an in-depth analysis of the pricing of thirteen real-world SaaS \cite{LABPACK}, we devised a common model that is able to represent any pricing of such systems in a unified way: \textbf{Pricing4SaaS} (see Figure \ref{fig:umlModel}), generalizing solutions like Pricing4API \cite{PRICING4API}.

Guided by the pricing model, within Pricing4SaaS, the user interacts with the pricing —encompassing plans, add-ons, or a combination thereof— through a subscription, which acts as a pivotal component in regulating the user's access to the IS functionalities according to the contract.



Both pricing components consist of \textbf{features}, which conform the minimum unit within the dissection of the structure. Depending on their contribution to the system, we have non-exhaustively devised up to eight different types, namely: 

\begin{itemize}
    \item \textbf{Domain:} provide functionality related to the domain of the \system, allowing to perform new operations or using exclusive services; e.g. \textit{voting in meetings}, which allows a meeting organizer to create single or multiple-choice survey questions within his meetings (Figure \ref{fig:zoomPricing}). This type of feature is usually the most frequent within a pricing.
    
    \item \textbf{Integration:} permit users to interact with the \system through its API, or to use functionalities from external third-party software within the \system; e.g. \textit{LTI Integration} on Zoom allows the addition of a meeting to any course activity created with a supported learning management system (LMS), such as Blackboard.

    \item \textbf{Automation:} they permit to configure the \system in order to perform some actions autonomously or track thresholds and notify the user if anyone is exceeded. It also includes any task performed by a bot or AI, such as predictions, generative AI, etc; e.g. Zoom's \textit{automated subtitles}.

    \item \textbf{Management:} are focused on team leaders and system administrators. They ease the supervision, organization and guidance of projects, and allow the configuration of accounts and organization-based restrictions and rules; e.g Zoom's \textit{Administrator portal}, which offers administrators a centralized tool to manage company's users for simplified billing and to declare advanced user management settings.

    \item \textbf{Information:} allow to see, use, visualize or extract additional data from the features described above; e.g. Zoom's \textit{reports}, which provides owners and account managers with a variety of account, meetings and webinars statistics.
    
    \item \textbf{Guarantee:} technical commitments of the company that operates the \system towards the users. For example, Zoom assures an \textit{end-to-end encryption} for meetings. 
    
    \item \textbf{Support:} expose the granularity of customer support offered within the plans; e.g. Zoom provides \textit{chat support} for PRO and BUSINESS plans.
    
    \item \textbf{Payment:} specify payment conditions and possibilities.
\end{itemize}

The model also allows to determine a second-level classification for features, that will be considered as \textbf{common} or \textbf{specific} whether they provide the same functionality across all plans or not, e.g \textit{automated subtitles} in Zoom is common. Such classification can be used to assess the level of privatization of a \SaaS.

Finally, features, add-ons and plans can be limited by \textbf{usage limits}, which are also present in APIs \cite{PRICING4API}. Depending on the limitation they impose, we have classified them into four groups:

\begin{itemize}
    \item \textbf{Renewable:} their limit is reset after a period of time, could be a day, week, month... For example, Zoom impose a \textit{max assistants per meeting} limit on its \textit{online meetings} feature.
    
    \item \textbf{Non-renewable:} define a static limit towards which the user approaches, and that will remain until the end of the subscription; e.g Zoom maximum \textit{cloud storage capacity} for recordings.
    
    \item \textbf{Response-driven:} represent a limit where user consumes more or less of his quota depending on the computational cost of the SaaS associated with the request. As an illustration, the usage of OpenAI's Large Language Models (LLMs)\footnote{OpenAI LLMs pricing is available at \href{https://openai.com/pricing}{URL}} are constrained by the number of tokens acquired. The consumption of tokens varies depending on the length of the model's response to a prompt.
    
    \item \textbf{Time-driven:} with this type the quota is consumed by usage time, and is normally combined with a non-renewable limit; e.g Zoom's \textit{max time per meeting}.
    
\end{itemize}

However, many SaaS offer add-ons to punctually extent these limits (similar to an API overage fee)\cite{GAMEZ2017}. For example, in Zoom, the maximum number of participants in a meeting can be extended by adding the \textit{huge meetings} add-on to the subscription, which allows up to 1000 participants in a meeting.

As can be seen, Pricing4SaaS abstracts from the underlying marketing strategy utilized to design a pricing, which enforces its flexibility.

\subsection{SaaS pricing validity criteria}

Defining a set of validation rules for pricings will complement Pricing4SaaS with additional benefits, e.g the automation of the validation process, which would allow the early detection of inconsistencies within the pricing before applying changes. This not only expedites the correction process but also prevents potential negative repercussions on customer experience and brand perception, economical losses, and so on. Furthermore, automated validation reduce the risk of human errors and facilitates the efficient management of variability. Thus, a valid pricing must verify the following conditions:

\begin{enumerate}
    \item There are not two plans with the same set of features and usage limits, as they must have differences among them by definition.
    
    \item Every feature contained within a pricing must be included in, at least, one plan or add-on, i.e there can be no published features that are inaccessible to users.
    
    \item A pricing can have as many limits as required, but they must verify that: i) the duration covered by a limit must not surpass the contractual period of the subscription; ii) they must have an associated objective metric through which the user can monitor the use of a feature.
\end{enumerate}

\subsection{Yaml4SaaS: a serialization of Pricing4SaaS}
\label{section:validation}

\begin{figure}
    \centering
    \makebox[\textwidth][c]{\includegraphics[width=\textwidth]{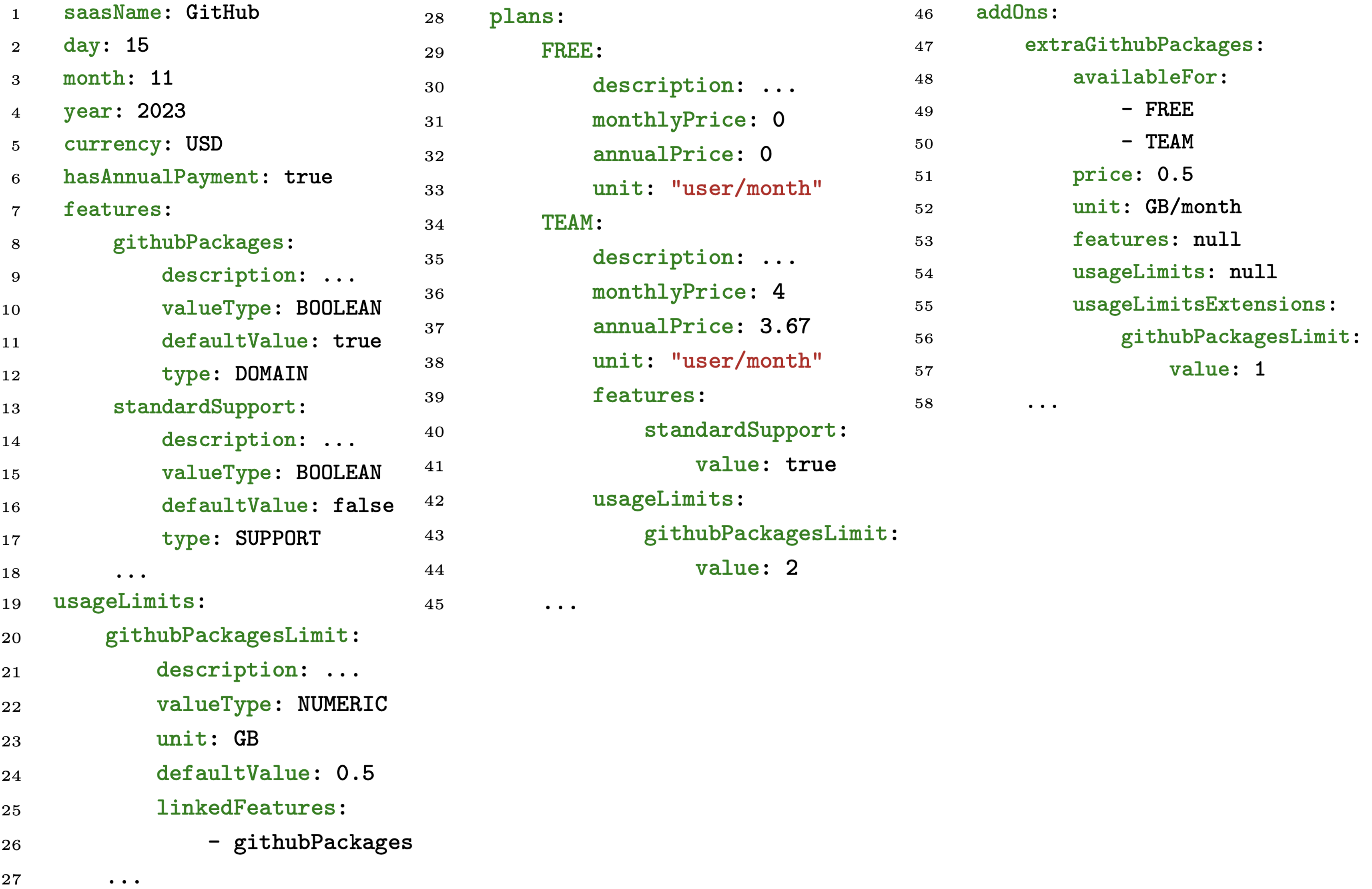}}
    \caption{Yaml4SaaS syntax} 
    \label{fig:yaml4SaaS}
\end{figure}

In order to demonstrate the applicability of Pricing4SaaS, we have designed a YAML-based syntax that describes a pricing with the directives of the model: \textbf{Yaml4SaaS}. This approach builds upon the accepted structure outlined in \cite{GAMEZ2019} and leverages the insights provided by Pricing4SaaS, enhancing the representation of pricing structures and establishing an unified and efficient way of capturing the intricacies of such pricing models (Figure \ref{fig:yaml4SaaS})\footnote{A complete explanation of the syntax can be found in \cite{LABPACK}}. By using this syntax, we were able to model up to 16 commercial SaaS (results available in \cite{YAML4SAASVALIDATOR}).

We also created a command-line validation tool (available in \cite{YAML4SAASVALIDATOR}) that checks Yaml4SaaS and performs the validations in the previous section. It is a part of a library we are currently developing: \textit{Pricing4Java}, a library built in Spring that offer a middleware component to intercepts the requests received by the back-end and evaluates the current state of the user subscription towards the pricing specification in Yaml4SaaS, stating which features are available under certain circumstances.

\section{Conclusions and future work}
\label{section:conclussions}

In this paper, \textit{Pricing-driven development and operation of SaaS} was presented as a conceptual framework that encapsulates the challenges and opportunities of managing SaaS systems whose pricing suffers frequent changes due to business factors. As an initial step towards developing automated tools that provide systems with self-adaptation to this changes, our research proposes Pricing4SaaS, a model designed to to represent pricings of SaaS, and which generalizes the already presented Pricing4API model \cite{PRICING4API}. In addition, as a complement to this model, we have developed a YAML-based syntax, coined as Yaml4SaaS, whose applicability have been demonstrated towards real-life software products.

Several challenges remain for future work, but the main target is to design a framework that elevates pricing schemes to a first-class consideration in the implementation of IS, requiring the minimum human intervention to apply changes to the system derived from these structures, such as: addition/removal of features, usage limits thresholds modification, etc. As the access regulation to exclusive features presented in pricing plans results in either enabling or disabling elements from the user interface (UI), which can be automated with feature toggling techniques \cite{FOWLER2023,jezequel2022feature,MAHDAVI2022}, we are developing a React library that automatically manages this type of toggling from a pricing serialized in Yaml4SaaS. On the other hand, we are focused in developing \textit{Pricing4Java} as a novel approach to pricing management in back-end services.

\vspace{-12pt}
\section*{Replicability \& verifiability}
\vspace{-6pt}
For the sake of replicability and verifiability, all the artifacts and datasets generated in this study are available in the laboratory package \cite{LABPACK}. This material comprises of the companion technical report, which supports pricing frequent changes statements, our serialization of the pricing of each specific SaaS using Yaml4SaaS, jupyter notebook used to extract the statistical results.
\vspace{-12pt}
\section*{Aknowledgements}
\vspace{-6pt}
Authors are thankful to Pedro Gonzalez Marcos for his support on the modeling of the pricings of \SaaS.
This work has been partially supported by grants PID2021-126227NB-C21 and PID2021-126227NB-C22 funded by MCIN / AEI / 10.13039 / 501100011033 / FEDER, UE, and grants TED2021-131023B-C21 and TED2021-131023B-C22 funded by MCIN / AEI / 10.13039 / 501100011033 and by European Union  “NextGenerationEU”/PRTR.
\vspace{-12pt}

\bibliographystyle{splncs04}
\bibliography{references}

\end{document}